\documentclass[a4paper,11pt]{article}
\usepackage{amsmath,amssymb}
\usepackage{array}
\usepackage{graphicx}

\newcommand{\inp}[1]{\langle #1 \rangle}
\newcommand{\mc}[1]{\mathcal #1}
\newcommand{\AdS}{AdS_5}
\newcommand{\Db}{\overline{D}}
\newcommand{\D}{\overline{\mathbf{D}}}
\newcommand{\p}{\partial}
\newcommand{\tr}{\operatorname{tr}}

\title{Four-point functions of different-weight operators in the AdS/CFT correspondence}
\author{Le\'on Berdichevsky\footnote{E-mail: leonberdichevsky@yahoo.com.mx}
\and Pieter Naaijkens\footnote{E-mail: pieter@naaijkens.nl}\\ \and
\\ \\
{\it Institute for Theoretical Physics and Spinoza Institute} \\
{\it Utrecht University, 3508 TD Utrecht, The Netherlands}}
\date{}
\begin{document}

\thispagestyle{empty}
\begin{titlepage}
\begin{flushright}
  ITP-UU-07/47 \\
  SPIN-07/35
\end{flushright}
  \renewcommand{\thefootnote}{\fnsymbol{footnote}}
  \makeatletter
  \let\footnotesize\small
  \let \footnote \thanks
  \null\vfil
  \vskip 40\p@
  \begin{center}%
    {\LARGE \@title \par}%
    \vskip 3em%
    {\large
     \lineskip .75em%
      \begin{tabular}[t]{c}%
        \@author
      \end{tabular}\par}%
      \vskip 1.5em%
    {\large \@date \par}
  \end{center}\par
  \@thanks
  \vfil\null
  \makeatother

\begin{abstract}
  We calculate four-point correlation functions of two weight-2 and two weight-3 $\frac{1}{2}$-BPS
  operators in $\mathcal{N}=4$ SYM in the large $N$ limit in supergravity approximation. By the AdS/CFT conjecture, these
  operators are dual to $\AdS$ supergravity scalar fields $s_2$ and $s_3$ with mass
  $m^2 = -4$ and $m^2 = -3 $ respectively. This is the first non-trivial four-point function of
  mixed-weight operators of lowest conformal dimensions.

  We show that the supergravity-induced four-point function splits into a ``free" and a ``quantum" part, where the quantum contribution obeys non-trivial constraints coming from the
  insertion procedure in the gauge theory, in particular, it depends on only
  \emph{one} function of the conformal cross-ratios. 
\end{abstract}
\vspace{\fill}
\end{titlepage}
\newpage

\setcounter{page}{1}
\renewcommand{\thefootnote}{\arabic{footnote}}
\setcounter{footnote}{0}

\section{Introduction}
Based on investigations of near horizon geometries and scattering
from black hole metrics, it was conjectured
\cite{Maldacena:1997re,Witten:1998qj,Gubser:1998bc} that the large
$N$ limit of some superconformal gauge theories in $d$-dimensional
flat space-time is governed by string (supergravity) theories on
$d+1$-dimensional anti-de Sitter space ($AdS_{d+1}$) times a
$10-(d+1)$ compact manifold ($\mathcal{M}^{10-(d+1)})$.\footnote{For
a review of the correspondence see e.g.\cite{Aharony:1999ti}.} In
particular, strongly coupled $\mathcal{N} = 4$ supersymmetric
Yang-Mills theory in four dimensions (SYM) is conjectured to be
dual to Type IIB supergravity on an $\AdS \times S^5$ background.

The compactification of Type IIB supergravity on $AdS_5\times S^5$
gives rise to an infinite tower of massive Kaluza-Klein (KK) modes
in the resulting five-dimensional theory. The isometry group of Type
IIB supergravity is identical to the superconformal group of the
dual field theory. The kinematical relation between the two theories
implies that the scalar KK modes $s_k \;(k \geq 2)$ that are
mixtures of the five-form potential and the graviton on $S^5$, are
dual to $\frac{1}{2}$-BPS operators of $\mathcal{N}=4$ SYM.  
These operators form short superconformal multiplets and have
conformal dimensions that are protected from quantum corrections. By
definition, the highest-weight operators of these multiplets are
annihilated by half of the Poincar\'e supercharges.

Another kinematical consequence of the identification mentioned
above is that two- and three-point functions of $\frac{1}{2}$-BPS
operators are protected from quantum corrections. Therefore, the
calculation of the three-point supergravity-induced correlators
doesn't give any new dynamical information and one needs to go
further to test the conjectured correspondence.

The four-point functions of such operators are not protected from
quantum corrections and therefore are the simplest non-trivial
candidates to explore the dynamics in the strong coupling limit.
Furthermore, the quantum behavior of four-point functions is
severely restricted, not only kinematically due to conformal
invariance, but also dynamically due to the existence of the
Lagrangian of $\mathcal{N} = 4$ SYM. The latter is given by the
insertion procedure~\cite{Intriligator:1998ig}, and reduces the
functional freedom predicted by conformal invariance. This inherent
dynamical feature of $\mathcal{N}=4$ SYM has no analogue in Type IIB
supergravity and offers the possibility of testing the AdS/CFT
correspondence by comparing the supergravity-induced results with
the structure predicted by it~\cite{Eden:2000bk}.

The five-dimensional effective action for Type IIB supergravity
relevant for the calculation of supergravity-induced four-point
functions of $\frac{1}{2}$-BPS operators has already been
constructed~\cite{Arutyunov:1999fb}. The most important feature of
the Lagrangian is that quartic couplings have at most four
derivatives in the fields. Due to the involved computational work
and complexity of the couplings, only specific examples of
four-point functions involving four identical operators have been
calculated, namely, for the operators with weight
$k=2,3,4$~\cite{Arutyunov:2000py,Arutyunov:2002fh,Arutyunov:2003ae}.\footnote{For
four-point correlators of non-superconformal primary operators,
see~\cite{Uruchurtu:2007kq,D'Hoker:1999pj,Liu:1998ty}.} It has been
found that the supergravity results indeed have the dynamical
structure predicted by $\mathcal{N}=4$ SYM.

In this paper we go beyond and explore the gauge/supergravity duality calculating the
first non-trivial four-point function of \emph{mixed}
$\frac{1}{2}$-BPS operators with lowest conformal dimensions,
\emph{i.e.}, the correlation function involving two $k=2$ and two
$k=3$ operators. We first establish the general structure
predicted by conformal invariance and the insertion procedure, and
then compare it with the one calculated from supergravity.

To this end, we first extract the relevant fourth order Lagrangian
from the general one. We find that the four-derivatives quartic
couplings can be reduced to two and non-derivatives couplings.  Hence,
the relevant Lagrangian is of  $\sigma$-\emph{model type}. This
characteristic was also found for the relevant Lagrangians necessary
for the computation of the four-point functions mentioned above. The
cancellation of four derivatives in the present case was indeed
expected since the relevant couplings are
sub-subextremal~\cite{Arutyunov:2000im}.

We again find that the amplitude splits into a ``free''
and a ``quantum'' part which exactly coincide with the result
calculated from $\mathcal{N}=4$ SYM and the prediction given by
the insertion procedure. Since in the supergravity side there is not
a quantity analogous to the coupling constant $g_{YM}$, the latter
can be interpreted as another non-trivial check supporting the AdS/CFT
correspondence.

The paper is organized as follows: in
Section~\ref{sec:correlationcft} the general form of the four-point
function under consideration is found from conformal, $R$- and
crossing symmetries. Further constraints on the coefficients are
found by the insertion procedure. In Section~\ref{sec:lagrangian},
we obtain the Lagrangian that we use to compute the
supergravity-induced amplitude. Finally, in Section~\ref{sec:concl}
the result obtained from the supergravity analysis is compared with
the conformal field theory predictions.  Technical details are
gathered in the Appendices, in particular, the $C$-algebra (where we
include the normalized projectors) and the novel computation of
vector and massive symmetric tensor exchange diagrams, where the
vector and tensor couple to currents which are not conserved on-shell.

\section{Generalities}
\label{sec:correlationcft}%
The conformal structure of $\textrm{SYM}$ restricts the form of the four-point function. In this
section the general form of the four-point function of two weight-2 and two weight-3
$\frac{1}{2}$-BPS operators is discussed. 

The $\frac{1}{2}$-BPS operators of conformal weight $k$ we
consider are single-trace operators in the $\mc{N}=4\;
\textrm{SYM}$ theory, given by\footnote{For $k \geq 4$ the
scalar fields $s_k$ correspond to \emph{extended}
$\frac{1}{2}$-BPS operators, where these single-trace operators
receive a multi-trace correction. For regular correlators however,
this operator mixing is suppressed in the large $N$
limit~\cite{Arutyunov:1999en,Arutyunov:2000im}.}
\begin{equation}
  \mathcal{O}_k^{I} = C_{i_1 \cdots i_k}^I \tr(\phi^{i_1} \cdots \phi^{i_k}).
  \label{eq:halfbps}
\end{equation}
The fields $\phi^i, i = 1, \dots 6$ are $\mc{N} = 4$ SYM scalar fields
and the $C^I_{i_1, \dots i_k}$ are traceless symmetric $SO(6)$
tensors (see Appendix~\ref{ap:ctensor}). The index $I$ runs over the
basis of the corresponding $SO(6)$ irrep with Dynkin labels
$[0,k,0]$. We want to find the general structure of the four-point
function
\[
\inp{\mc{O}_2^{I_1}\mc{O}_2^{I_2}\mc{O}_3^{I_3}\mc{O}_3^{I_4}} \equiv
\inp{\mc{O}_2^{1}\mc{O}_2^{2}\mc{O}_3^{3}\mc{O}_3^{4}}
\]
and compare it with the result we obtain in the supergravity approximation.

\begin{figure}
  \begin{center}
    \includegraphics[scale=0.8]{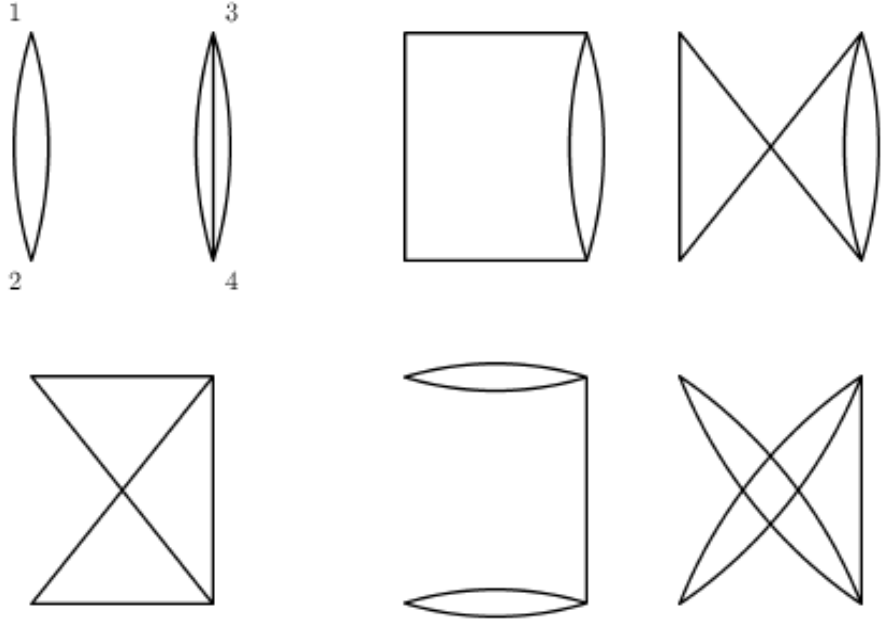}%
  \end{center}
  \caption{Propagator structures. The diagrams are divided into
four equivalence classes. The elements in an equivalence class are related by crossing symmetry.}
  \label{fig:contract4pt}
\end{figure}
We will apply the methods of Sections 2 and 3 of~\cite{Arutyunov:2002fh} (see also references there)
to our specific case. This gives a basis in terms of the tensor structures appearing. There are six
different structures belonging to four equivalence classes, which are shown in
Figure~\ref{fig:contract4pt}. They are called propagator structures, since they appear naturally by
connecting propagators of the scalar fields $\phi^i$. The elements in an equivalence class are
related by crossing symmetry.  To get the most general conformally invariant form of the correlator,
we multiply each structure by a function of the conformal cross-ratios
\begin{equation}
  s = \frac{x_{12}^2 x_{34}^2}{x_{13}^2 x_{24}^2},\quad  t = \frac{x_{14}^2 x_{23}^2}{x_{13}^2
  x_{24}^2},
  \label{eq:crossratios}
\end{equation}
where $x_{ij}^2 = |\vec{x}_i - \vec{x}_j|^2$. Following this procedure gives, in terms of the
propagator basis from Appendix~\ref{sec:calgebra},
\begin{equation}
  \begin{split}
    \inp{O_2^1(x_1) &O_2^2(x_2) O_3^3(x_3) O_3^4(x_4)} = a(s,t) \frac{\delta_2^{12}
    \delta_3^{34}}{x_{12}^4 x_{34}^6} + b(s,t) \frac{S^{1234}}{x_{12}^2 x_{13}^2 x_{14}^2 x_{23}^2
    x_{24}^2}
    \\ &+ c_1(s,t) \frac{C^{1243}}{x_{12}^2 x_{34}^4 x_{13}^2 x_{24}^2} +
    c_2(s,t) \frac{C^{1234}}{x_{12}^2 x_{34}^4 x_{14}^2 x_{23}^2} \\
    &+ d_1(s,t) \frac{\Upsilon^{1234}}{x_{13}^4 x_{34}^2 x_{24}^4 } + d_2(s,t)
    \frac{\Upsilon^{1243}}{x_{34}^2 x_{23}^4 x_{14}^4}.
  \end{split}
  \label{eq:4pointcft}
\end{equation}
Here, the $O_k$ denote the canonically normalized versions of the operators $\mathcal{O}_k$, such
that $\inp{O_k^{I_1}(x_1) O_k^{I_2}(x_2)} = \frac{\delta^{I_1 I_2}}{x_{12}^{2k}}$.  This is the most
general form, allowed by conformal- and R-symmetry, of the four-point function we consider.

By simple symmetry considerations, relations between the coefficient
functions can be derived. Note that when permuting $x_1
\leftrightarrow x_2$, the cross-ratios transform as $s \to s/t$, and
$t \to 1/t$. On the other hand, in the RHS of
eq.\eqref{eq:4pointcft}, this corresponds to interchanging the
representation labels 1 and 2. Using the symmetry properties of the
$C$-tensors, it then immediately follows that
\begin{equation}
\begin{split}
 a(s,t) = a(s/t, 1/t), \quad b(s,t) = b(s/t, 1/t), \\
 \quad c_1(s/t,1/t) = c_2(s, t), \quad d_1(s/t, 1/t) = d_2(s,t).
 \end{split}
 \label{eq:crossingsym}
\end{equation}
These are the crossing symmetry relations.

There is another relation between these coefficient functions. This
is based on the insertion formula~\cite{Intriligator:1998ig}. This
procedure gives additional constraints on the \emph{quantum} part of
the correlator. From the results of Section 3 of
Ref.\cite{Arutyunov:2002fh}, we find that the coefficient functions
can be expressed in terms of a \emph{single} function $\mc{F}(s,t)$ by
\begin{equation}
a(s,t) = s \mc{F}(s,t), \quad  d_1(s,t) = \mc{F}(s,t), \quad d_2(s,t) = t \mc{F}(s,t),
\label{eq:insert1}
\end{equation}
and
\begin{equation}
  \begin{split}
    b(s,t) &= (s-t-1) \mc{F}(s,t), \\
    c_1(s,t) &= (t-s-1) \mc{F}(s,t), \\
    c_2(s,t) &= (1-s-t) \mc{F}(s,t).
  \end{split}
  \label{eq:insert2}
\end{equation}
The crossing symmetry relations~\eqref{eq:crossingsym} then imply
that $\mc{F}(s/t, 1/t) = t \mc{F}(s,t)$. Hence, all dynamical
information in the four-point function is completely determined by
the single function $\mc{F}(s,t)$ of the cross-ratios.  This result
is purely based on conformal field theory considerations and the
insertion procedure.  Later we will compare the result from
supergravity calculations with this general form.

Finally, the coefficient functions in free field theory are, in the large $N$ limit, given by
\begin{equation}
  a = 1, \quad b = \frac{12}{N^2}, \quad c_{1,2} = \frac{6}{N^2}, \quad d_{1,2} =
  0.
  \label{eq:freefield}
\end{equation}
Note that the color structure of the diagrams implies that the free field contribution from
\emph{one-particle reducible} diagrams identically vanishes.

\section{Supergravity Lagrangian}\label{sec:lagrangian}
The computation of four-point functions of $\frac{1}{2}$-BPS operators in supergravity aproximation
requires the $5d$ effective quartic action of compactified Type IIB supergravity on an $AdS_5 \times
S^5$ background, and the identification of  the relevant parts. The effective five-dimensional
action can be written as~\cite{Arutyunov:1999fb}
\begin{equation}
  S = \frac{N^2}{8 \pi ^2} \int \textrm{d}^5 z \sqrt{g} ( \mathcal{L}_2 + \mathcal{L}_3 +
  \mathcal{L}_4),
  \label{eq:efaction}
\end{equation}
a sum of quadratic, cubic and quartic terms. In comparison with Ref.~\cite{Arutyunov:1999fb}, we
will work on the Euclidean version of $AdS_5$, which results in an overall minus sign, and rescale
the fields to match the quadratic part presented below. In equation~\eqref{eq:efaction}, $g$ is the
determinant of the Euclidean metric on $AdS_5$, $\textrm{d} s^2 = \frac{1}{z_0^2}(\textrm{d}z_0^2 +
\textrm{d} x^i \textrm{d} x^i)$, with $i=1,\dots,4$.

The quadratic terms were found in~\cite{Arutyunov:1998hf}. The relevant part in this case is
\begin{equation}
  \begin{split}
  \mc{L}_2 = &\frac{1}{4}\left( \nabla_\mu s_2^1 \nabla^\mu s_2^1 - 4 s_2^1 s_2^1 \right) +
  \frac{1}{4}\left( \nabla_\mu s_3^1 \nabla^\mu s_3^1 - 3 s_3^1 s_3^1 \right) \\ &+ \frac{1}{2}
  (F_{\mu\nu,1}^1)^2 + \frac{1}{2} (F_{\mu\nu,2}^1)^2 + 3 (A_{\mu,2}^1)^2 \\ &+ \frac{1}{4}
  \nabla_\rho \phi_{\mu\nu,0} \nabla^\rho \phi_0^{\mu\nu} - \frac{1}{2} \nabla_\mu
  \phi_{\mu\rho,0}\nabla^\nu \phi_{\nu\rho,0} + \frac{1}{2} \nabla_\mu \phi^\rho_{\rho,0} \nabla_\nu
  \phi^{\mu\nu}_0 - \frac{1}{4} \nabla_\rho \phi_{\mu,0}^\mu \nabla^\rho \phi^\nu_{\nu,0} \\ &-
  \frac{1}{2} \phi_{\mu\nu,0} \phi^{\mu\nu}_0 + \frac{1}{2}(\phi_{\nu,0}^\nu)^2 \\
  &+ \frac{1}{4} \nabla_\rho \phi_{\mu\nu,1}^1 \nabla^\rho \phi_1^{\mu\nu\,1} - \frac{1}{2} \nabla_\mu
  \phi_{\mu\rho,1}^1 \nabla^\nu \phi_{\nu\rho,1}^1 + \frac{1}{2} \nabla_\mu \phi^{\rho\,1}_{\rho,1} \nabla_\nu
  \phi^{\mu\nu\,1}_1 - \frac{1}{4} \nabla_\rho \phi_{\mu,1}^{\mu\,1} \nabla^\rho \phi^{\nu\,1}_{\nu,1} \\
  &+ \frac{3}{4} \phi_{\mu\nu,1}^1 \phi^{\mu\nu\,1}_1 - \frac{7}{4}(\phi_{\nu,1}^{\nu\,1})^2,
  \end{split}
\end{equation}
where $F_{\mu\nu,k} = \partial_\mu A_{\nu,k} - \partial_\nu A_{\mu,k}$, and a summation over the
upper indices, which run over the basis of the irrep corresponding to the field, is
implied. A novelty in these calculations is the appearance of the vector $A_{\mu,1}$ and the tensor
$\phi_{\mu\nu,1}$. These appear coupled to one $s_2$ and one $s_3$
scalar field, which lead to tree diagrams where these fields are
exchanged. In previous calculations, these fields were not present
because of $SO(6)$ selection rules.

The cubic couplings needed to calculate four-point functions of arbitrary $\frac{1}{2}$-BPS
operators were calculated in Refs.~\cite{Lee:1998bx,Arutyunov:1999en,Lee:1999pj}. The couplings are given
in terms of $C$-tensors, which are related to spherical harmonics on $S^5$. They are the
Clebsch-Gordan coefficients for tensor products of two SO(6) irreps. For definitions and normalization,
we refer to Appendix B of Ref.~\cite{Arutyunov:2002fh}. We will employ the notation
\[
\inp{C^1_{k_1} C^2_{k_2} C^3_{[a_1, a_2, a_3]}}
\]
where lower indices $k_i$ denote the irrep with Dynkin labels $[0,k_i,0]$, and the upper indices run over the
basis of the corresponding irrep. With this notation the cubic couplings become
\begin{equation}
  \begin{split}
\mc{L}_3 = & -\frac{1}{3}\inp{C^1_2 C^2_2 C^3_{[0,2,0]}} s_2^1 s_2^2 s_2^3 -3 \inp{C^1_3 C^2_3 C^3_{[0,2,0]}} s_3^1
  s_3^2 s_2^3 \\
  & -\frac{1}{4}\left(\nabla^\mu s_2^1 \nabla^\nu s_2^1
  \phi_{\mu\nu,0} - \frac{1}{2} \left( \nabla^\mu s_2^1 \nabla_\nu s_2^1 - 4 s_2^1 s_2^1 \right)
  \phi^\nu_{\nu,0} \right) \\
  & -\frac{1}{4}\left(\nabla^\mu s_3^1 \nabla^\nu s_3^1
  \phi_{\mu\nu,0} - \frac{1}{2} \left( \nabla^\mu s_3^1 \nabla_\nu s_3^1 - 3 s_3^1 s_3^1 \right)
  \phi^\nu_{\nu,0} \right) \\
  &-\frac{1}{2} \inp{C^1_2 C^2_3 C^3_{[0,1,0]}} \left(\nabla^\mu s_2^1
  \nabla^\nu s_3^2 \phi_{\mu\nu,1}^3 + \frac{1}{2} \left( \nabla^\mu
  s_2^1 \nabla_\mu s_3^2 -6 s_2^1
  s_3^2 \right) \phi^{\nu 3}_{\nu,1} \right) \\
  &- \inp{C^1_2 C^2_2 C_{[1,0,1]}} s_2^1 \nabla^\mu
  s_2^2 A_{\mu,1}^3 - \frac{3}{2} \inp{C^1_3 C^2_3 C^3_{[1,0,1]}} s_3^1 \nabla^\mu s_3^2 A_{\mu,1}^3 \\
  &-\sqrt{3}\inp{C^1_2 C^2_3 C^3_{[1,1,1]}} s_2^1
  \nabla^\mu s_3^2 A^3_{\mu,2} -\sqrt{3}\inp{C^1_3 C^2_2 C^3_{[1,1,1]}} s_3^1
  \nabla^\mu s_2^2 A^3_{\mu,2}.
\end{split}
  \label{eq:cubic}
\end{equation}
One interesting feature one can read off from the Lagrangian is the
appearance of exchange diagrams of a vector and a symmetric tensor
fields which do not couple to conserved currents: the weight of the
external scalar fields are different.

\begin{figure}
  \begin{center}
    \includegraphics[scale=0.8]{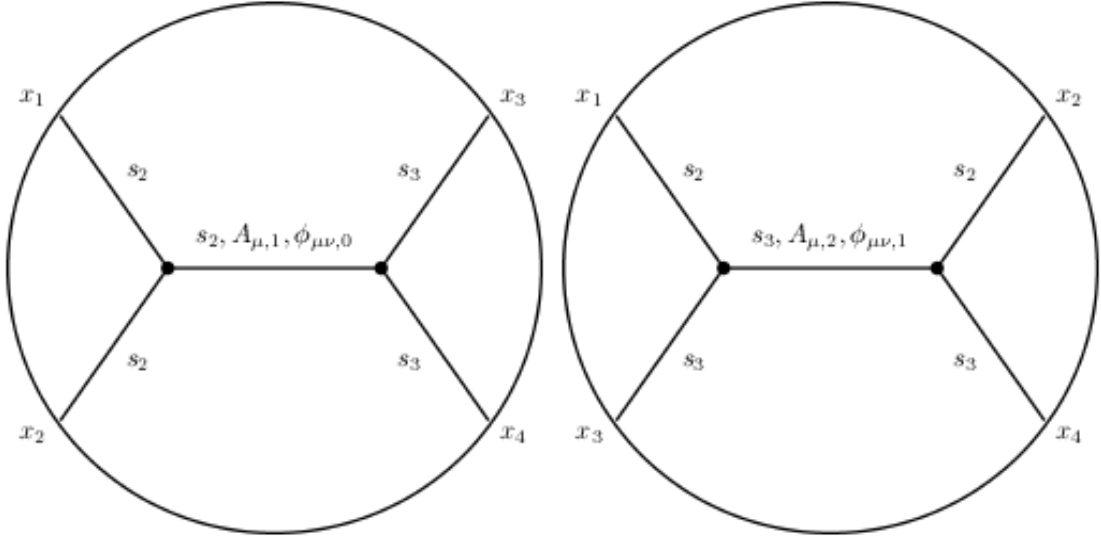}
  \end{center}
  \caption{Witten diagrams contributing to the four-point function.}
  \label{fig:exchange}
\end{figure}

The quartic couplings are the hardest to compute. The result (the details are in
Appendix~\ref{sec:quartic}) is
\[
  \begin{split}
  \mathcal{L}_4 = -\frac{1}{4}&\left( C^{1234} - S^{1234} \right) s_2^1 \nabla_\mu s_2^2 s_3^3 \nabla^\mu
  s_3^4 \\ &+ \frac{3}{8}\left( 9 C^{1234} + 5 S^{1234} - \delta_2^{12} \delta_3^{34} - 3 \Upsilon^{1234}
  \right) s_2^1 s_2^2 s_3^3 s_3^4.
  \end{split}
\]
Now that the Lagrangian have been found, we can determine its on-shell value. This amounts to calculating
exchange (and contact) diagrams. Witten diagrams of all exchange integrals
contributing are shown in Figure \ref{fig:exchange}. In comparison with previous results, we also have to compute the
exchange of a massive vector and a massive tensor coupled to currents which aren't conserved
on-shell. The method we use to calculate them is described in Appendix~\ref{sec:exchange}.

\section{Verifying CFT predictions and conclusions}\label{sec:concl}
After the on-shell value of the Lagrangian is calculated, we can trivially find the four-point
function. After finding the coefficient functions in terms of $D$-functions, we express them in
terms of the cross-ratios $s$ and $t$ by means of the $\Db$-functions (see Appendix~\ref{ap:dfunc}).
We find that the four-point function has indeed the structure of eq.\eqref{eq:4pointcft}.

There are two ways to verify the relations predicted by the insertion procedure. First of all, one
could work with the $\Db$-functions, and use the identities in Appendix D.2
of~\cite{Arutyunov:2002fh}. Crossing symmetry and other relations between $\Db$-functions are listed there. An advantage of this method is that it is possible to obtain a simple form of the
coefficient functions. In the present situation, however, the above method becomes rather cumbersome and error-prone, due to
the large number of different $\Db$-functions involved. A more straightforward method is to express
$\Db$-functions in terms of differential operators $\D$, defined in Appendix~\ref{ap:dfunc}. Using
equations~\eqref{eq:boxpartial}, it is then possible to write the coefficients completely in terms
of $\ln s, \ln t$ and $\Phi(s,t)$. Using that $\Phi(s/t, 1/t) = t \Phi(s,t)$, one finds that the
crossing symmetries~\eqref{eq:crossingsym} are indeed satisfied. This however is a rather trivial
check, as these crossing symmetries follow automatically if one considers the correct permutations
when calculating the individual contributions to the four-point function.

To check the insertion procedure predictions, we read off the single function $\mc{F}(s,t)$ from the
coefficient $a(s,t)$:\footnote{The reason to read it off from $a(s,t)$ is that there is no free field
contribution to the \emph{connected} four-point function for the $\delta_{2}^{12}\delta_{3}^{34}$
tensor.}
\[
\mc{F}(s,t) = \frac{3}{N^2} \left(-\Db_{1133}(s,t) + 4s \Db_{2233}(s,t) + (1-s+t) \Db_{2244}(s,t) \right),
\]
where $a(s,t)=s\mc{F}(s,t)$.

Then, by using the second method described above, we find
\begin{equation}
  \begin{split}
    b(s,t) - (s-t-1) \mc{F}(s,t) & = \frac{12}{N^2} \\
    c_1(s,t) - (t-s-1) \mc{F}(s,t) & = \frac{6}{N^2} \\
    d_1(s,t) - \mc{F}(s,t) & = 0,
  \end{split}
  \label{eq:ffcoeff}
\end{equation}
and similar for $c_2$ and $d_2$. The comparison of these results
with the restrictions for the quantum parts predicted by
eqs.\eqref{eq:insert1} and~\eqref{eq:insert2} shows that these are
indeed satisfied by the supergravity-induced four-point function, up
to some constants. But for $c_1, c_2, d_1, d_2$ and $b$ these
constants are precisely the free field contributions
eq.\eqref{eq:freefield}! Hence, we conclude that the coefficients
split into a quantum and a free field part, where the latter is due to \emph{one-particle irreducible} (1PI) diagrams.

The novelty of this result is that it is the first time four-point
correlators of $\frac{1}{2}$-BPS operators of different weight
(apart from some trivial (extremal) cases) have been calculated from
Type IIB supergravity. However, we do find the same features as for the case
where all operators are of equal weight $k=2,3,4$; namely, that the quantum part of the supergravity-induced amplitude satisfies the constrains obtained from the insertion
procedure in the field theory side, reducing the functional freedom to one (or two in the
case $k=4$) function of the conformal cross-ratios. This gives
further support for the AdS/CFT correspondence, since there is no
obvious explanation on the supergravity side for these results.

It would be interesting to compare our findings with the ones
obtained from perturbative $\textrm{SYM}$, since both
results correspond to a different regime. Furthermore, one can find
the OPE of operators involved in this correlator, to compute
anomalous dimensions~\cite{Arutyunov:2000ku,Dolan:2001tt,Dolan:2003hv,Dolan:2006ec}.

Another interesting question is to understand better the implications of the insertion procedure in
CFT on the supergravity side. These considerations could possibly lead to a simpler description of
the $\AdS$ supergravity. One could also include corrections to the supergravity result, for example
by considering D-instanton effects.
\\ \\
\textbf{Acknowledgements:} We would like to thank G. Arutyunov for
helpful discussions. We are also grateful to H. Osborn for pointing out misprints in
 the manuscript and O.~Aharony for pointing out the vanishing of the one-particle reducible diagrams in free field theory.

\appendix
\section{C-Algebra}
\label{sec:calgebra}\label{ap:ctensor}
The couplings of the five-dimensional action are described in terms of integrals of spherical
harmonics on $S^5$. These spherical harmonics can be described by
$C$-tensors $C^I_{i_1\dots i_k}$. The $C$-tensor transforms in the $[0,k,0]$ irrep of SO(6) (see the Appendices of
\cite{Lee:1998bx,Arutyunov:1999en}). These $C$-tensors are symmetrized according to the
corresponding Young pattern of the irrep. they correspond to. The integrals of the spherical
harmonics appearing in the couplings are basically the Clebsch-Gordon coefficients of products of two
SO(6) irreps, and can be expressed in terms of SO(6) rank 3 tensors
\[
\inp{C_{k_1}^1 C_{k_2}^2 C_{[a_1, a_2, a_3]}^3},
\]
where we denote the representation index $I_1$ by 1, etc. These rank
3 tensors are constructed by contracting particular subset of
indices of the $C$-tensors $C^{I_1}_{[0,k_1,0]}$,
$C^{I_2}_{[0,k_2,0]}$ and $C^{I_3}_{[a_1,a_2,a_3]}$. See Appendix B
of Ref.~\cite{Arutyunov:2002fh} for details.

The quartic couplings are described in terms of products of two Clebsch-Gordan coefficients, that arise from overlapping integrals of spherical harmonics
\begin{equation}
\inp{C_{k_1}^1 C_{k_2}^2 C_{[a_1, a_2, a_3]}^5}\inp{C_{k_3}^3 C_{k_4}^4 C_{[a_1, a_2, a_3]}^5}.
\label{eq:clebschsum}
\end{equation}
Since there is a summation over the representation index $I_5$, we can use a completeness relation
to express this tensor in the so-called propagator basis. In these calculations, the expressions
in the appendix of Ref.~\cite{Arutyunov:1999fb} are helpful. One can refer
to~\cite{Arutyunov:2002fh} for an example of their application.

The procecudure outlined above yields four different kinds of rank 4 tensors. One should notice that
for each kind, there are two $C$-tensors transforming as $[0,2,0]$, and two as $[0,3,0]$, involved.
This should be taken into account carefully when permuting the representation indices. The tensor
structures involved are given by
\begin{eqnarray}
  \label{eq:tstrucdelta}
  \delta^{12}_2 \delta^{34}_3 &= C^1_{ij} C^2_{ij} C^3_{klm} C^4_{klm}, \\
  C^{1234} &= C^1_{ij} C^2_{jk} C^3_{klm} C^4_{ilm},\\
  \Upsilon^{1234} &= C^1_{ij} C^2_{lm} C^3_{ijk}  C^4_{lmk}, \\
  S^{1234} &= C^1_{ik} C^2_{jl} C^3_{lkm} C^4_{ijm}.
  \label{eq:tensorstruc}
\end{eqnarray}
The tensors $\delta^{12}_2 \delta^{34}_3$ and $S^{1234}$ are symmetric under $1 \leftrightarrow 2$
and $3 \leftrightarrow 4$ separately, while $C^{1234}$ and $\Upsilon^{1234}$ have the following
symmetry relations:
\begin{equation}
  C^{1234} = C^{2143},\quad \Upsilon^{1234} = \Upsilon^{2143}.
  \label{eq:symtensor}
\end{equation}

We now evaluate eq.\eqref{eq:clebschsum} for the cases of interest for us. In the case of $k_1=k_2=2,
k_3=k_4=3$, this gives the following results. Note that the
selection rules for $k_5$ allow only for these values of $k_5$.
\begin{equation}
  \begin{split}
  \inp{C^1_2 C^2_2 C^5_{[0,0,0]}}\inp{C^3_3 C^4_3 C^5_{[0,0,0]}} &= \delta_2^{12} \delta_3^{34}, \\
  \inp{C^1_2 C^2_2 C^5_{[0,2,0]}}\inp{C^3_3 C^4_3 C^5_{[0,2,0]}} &= \frac{1}{2} C^{1234} +
  \frac{1}{2} C^{1243} - \frac{1}{6} \delta_{2}^{12} \delta_{3}^{34}, \\
  \inp{C^1_2 C^2_2 C^5_{[0,4,0]}}\inp{C^3_3 C^4_3 C^5_{[0,4,0]}} &= -\frac{2}{15} C^{1234}
  -\frac{2}{15}C^{1243} + \frac{2}{3} S^{1234} \\
  & \quad +\frac{1}{6} \Upsilon^{1243} + \frac{1}{6} \Upsilon^{1234} +\frac{1}{60} \delta_2^{12} \delta_3^{34}.
\end{split}
\label{eq:asum2233}
\end{equation}
It is interesting to see that they are of the same form as those found in the cases where
$C^1,C^2,C^3$ and $C^4$ are all in the same representation~\cite{Arutyunov:2003ae,Arutyunov:2002fh},
if one makes the proper identifications.

For the summation over the vector representation we get
\begin{equation}
  \begin{split}
    \inp{C^1_2 C^2_2 C^5_{[1,0,1]}}\inp{C^3_3 C^4_3 C^5_{[1,0,1]}} &= 2(C^{1243}-C^{1234}), \\
    \inp{C^1_2 C^2_2 C^5_{[1,2,1]}}\inp{C^3_3 C^4_3 C^5_{[1,2,1]}} &= \frac{1}{3}(C^{1234}-C^{1243})
    + \frac{2}{3}(\Upsilon^{1234} - \Upsilon^{1243}).
  \end{split}
  \label{eq:tsum2233}
\end{equation}

And the tensor representation
\begin{equation}
  \begin{split}
    \inp{C^1_2 C^2_2 C^5_{[2,0,2]}}\inp{C^3_3 C^4_3 C^5_{[2,0,2]}} &= -\frac{2}{3}\left(C^{1234} +
    C^{1243}\right) + \frac{4}{3}\left(\Upsilon^{1234} + \Upsilon^{1243}\right) \\
    & \quad -\frac{8}{3} S^{1234} + \frac{2}{15} \delta_2^{12} \delta_3^{34}.
  \end{split}
\end{equation}

Next we consider the case where $k_1 = k_3 = 2$ and $k_2 = k_4 = 3$. These are distinctively
different from the cases encountered in previous work, since the selection rules now give other
values for the representation that is summed over. For summation over the scalar representation the
results are
\begin{equation}
  \begin{split}
    \inp{C^1_2 C^2_3 C^5_{[0,1,0]}}\inp{C^3_2 C^4_3 C^5_{[0,1,0]}} &= \Upsilon^{1324}, \\
    \inp{C^1_2 C^2_3 C^5_{[0,3,0]}}\inp{C^3_2 C^4_3 C^5_{[0,3,0]}} &= \frac{1}{3}C^{1342} +
    \frac{2}{3} S^{1324} - \frac{1}{6} \Upsilon^{1324}, \\
    \inp{C^1_2 C^2_3 C^5_{[0,5,0]}}\inp{C^3_2 C^4_3 C^5_{[0,5,0]}} &= \frac{3}{5}C^{1324}
    -\frac{1}{10} C^{1342} - \frac{1}{5} S^{1324} \\
    & \quad + \frac{1}{10}\delta_2^{13} \delta_3^{24} + \frac{3}{10} \Upsilon^{1342} +
    \frac{1}{50}\Upsilon^{1324},
  \end{split}
\end{equation}
and for the vector cases
\begin{equation}
  \begin{split}
    \inp{C^1_2 C^2_3 C^5_{[1,1,1]}}\inp{C^3_2 C^4_3 C^5_{[1,1,1]}} &= \frac{3}{2} C^{1342} -
    \frac{3}{2} S^{1324} - \frac{3}{10} \Upsilon^{1324}, \\
    \inp{C^1_2 C^2_3 C^5_{[1,3,1]}}\inp{C^3_2 C^4_3 C^5_{[1,3,1]}} &= \frac{5}{12} C^{1324} -
    \frac{25}{84} C^{1324} + \frac{5}{21} S^{1324} \\
    & \quad + \frac{5}{12} \delta_2^{13} \delta_3^{24} + \frac{1}{42} \Upsilon^{1324} -
    \frac{5}{6} \Upsilon^{1342}.
  \end{split}
  \label{eq:tsum2323}
\end{equation}
Finally, the tensor case gives
\begin{equation}
    \begin{split}
      \inp{C^1_2 C^2_3 C^5_{[2,1,2]}}\inp{C^3_2 C^4_3 C^5_{[2,1,2]}} &= -\frac{16}{9} C^{1324} -
      \frac{40}{63} C^{1342} - \frac{16}{63} S^{1324} \\
      & \quad + \frac{8}{9} \delta_2^{13} \delta_3^{24} + \frac{8}{63} \Upsilon^{1324} + \frac{8}{9}
      \Upsilon^{1342}.
    \end{split}
\end{equation}
The results of the remaining cases are the same, if one changes the representation labels
accordingly, except for equation~\eqref{eq:tsum2323}, which acquires an additional minus sign in the
cases $k_1=k_4=3, k_2=k_3=2$ and $k_1=k_4=2, k_2=k_3=3$.

\begin{table}
  \begin{center}
    \setlength{\extrarowheight}{7pt}
  \begin{tabular}{l|c c c c c c}
    Tensor & $\delta_2^{12} \delta_3^{34}$ & $C^{1234}$ & $C^{1243}$ & $\Upsilon^{1234}$ &
    $\Upsilon^{1243}$ & $S^{1234}$ \\
    \hline 
    $\delta_2^{12} \delta_3^{34}$ & $1000$  & $\frac{500}{3}$ & $\frac{500}{3}$ &
	    $50$ & $50$ & $\frac{50}{3}$ \\
    $C^{1234}$ & $\frac{500}{3}$  & $\frac{725}{3}$ & $\frac{25}{3}$ &
	    $\frac{25}{3}$ & $125$ & $\frac{25}{2}$ \\
    $\Upsilon^{1234}$ & $50$  & $\frac{25}{3}$ & $125$ &
    $\frac{1250}{3}$ & $\frac{25}{3}$ & $\frac{125}{3}$ \\
    $S^{1234}$ & $\frac{50}{3}$ & $\frac{25}{2}$ & $\frac{25}{2}$ & $\frac{125}{3}$ &
    $\frac{125}{3}$ & $\frac{425}{4}$
  \end{tabular}
\end{center}
\caption{Pairings between independent tensors}
\label{tb:pairings}
\end{table}
Finally, for completeness and in order to facilitate a future
analysis of the OPE of the four-point function, we include a table
of pairings between the elements of the propagator basis. These can
be found using the completeness relation. Our results are summarized
in Table~\ref{tb:pairings}.

With these pairings we can fix the normalization for the projectors
$P^{1234}_{[k_1, k_2, k_3]}$ to be $P^{1234}_{[k_1, k_2, k_3]}
P^{1234}_{[k_1, k_2, k_3]} = \textrm{dim} [k_1,k_2,k_3]$, and check
that they are orthogonal. These projectors can be used to project
the four-point correlator onto the irreps appearing in the tensor
decompositions of $[0,2,0] \times [0,3,0]$ and $[0,2,0] \times
[0,2,0]$. The projectors itself we have already found: they are
proportional to the summations of $C$-tensors we found above. For
example, $P^{1234}_{[0,1,0]} \propto \inp{C_2^1 C_3^3 C_{[0,1,0]}^5}
\inp{C_2^2 C_3^4 C_{[0,1,0]}^5}$. With Table~\ref{tb:pairings} it is
trivial to normalize them properly and check orthogonality.

\section{Reduction of quartic couplings} \label{sec:quartic}
In this section we describe how we can rewrite the four-derivative terms in the Lagrangian, and
end up with a remarkably simple expression for the quartic contribution to the Lagrangian. We show
that the Lagrangian is of \emph{$\sigma$-model type}, a feature that was also found for the
effective Lagrangian in calculations of equal-weight $k=2,3$ and $4$ four-point
functions~\cite{Arutyunov:2000py,Arutyunov:2002fh,Arutyunov:2003ae}. The fact that the Lagrangian
reduces to a simple expression in each case, suggests that there may be a simpler description.

The quartic couplings of the scalar fields are given in~\cite{Arutyunov:1999fb}. They can be written
as
\begin{equation}
\begin{split}
\mc{L}^4 = \mc{L}^{(4) I_1 I_2 I_3 I_4}_{k_1 k_2 k_3 k_4} s_{k_1}^{I_1} \nabla_\mu s_{k_2}^{I_2}
\nabla^\nu \nabla_\nu (s_{k_3}^{I_3} \nabla^\mu s_{k_4}^{I_4}) \\ +
\mc{L}^{(2) I_1 I_2 I_3 I_4}_{k_1 k_2 k_3 k_4} s_{k_1}^{I_1} \nabla_\mu s_{k_2}^{I_2}
s_{k_3}^{I_3} \nabla^\mu s_{k_4}^{I_4}+
\mc{L}^{(0) I_1 I_2 I_3 I_4}_{k_1 k_2 k_3 k_4} s_{k_1}^{I_1} s_{k_2}^{I_2} s_{k_3}^{I_3}
s_{k_4}^{I_4},
\end{split}
\end{equation}
where in the case we are interested in, two of the $k_i$'s are equal to 2, and the other two are equal
to 3. Hence, this allows for 6 possible permutations. The indices $I_i$ run over the basis of the
representation $[0,k_i,0]$, and should be summed over. From now on, we will denote this indices
simply as superscript 1,2,3 and 4.

We now proceed as in Refs.~\cite{Arutyunov:2002fh,Arutyunov:2003ae}. We re-expand the products Clebsch-Gordon
coefficients in the couplings, which form the so-called ``OPE basis'', given by
\[
\inp{C^1_{k_1} C^2_{k_2} C^5_{[a_1,a_2, a_3]}}\inp{C^3_{k_3} C^4_{k_4} C^5_{[a_1,a_2, a_3]}},
\]
over the propagator basis. This is done in Appendix~\ref{sec:calgebra}. The subscripts $k_i$ denote
that the $C$-tensor transforms according to the $[0,k_i,0]$ irrep.

\subsection{Four-derivative couplings}
In the four-derivative couplings we encounter two basic tensor structures. We will denote them as
\begin{equation}
  \begin{split}
    A^{1234} &= \frac{3}{5 \cdot 2^{16}} \left( 41 (C^{1234} - C^{1243}) + 31 (\Upsilon^{1234} -
    \Upsilon^{1243}) \right)\\
    \Sigma^{1234} &= \frac{9}{5 \cdot 2^{18}}  \left(6 (C^{1234} + C^{1243}) + 12 S^{1234} +
    \delta_2^{12}\delta_3^{34} + 3 (\Upsilon^{1234} + \Upsilon^{1243}) \right).
  \end{split}
  \label{eq:tensor4}
\end{equation}
Note that $A^{1234}$ is anti-symmetric under $3 \leftrightarrow 4$, while $\Sigma^{1234}$ is
symmetric. With this notation, the four-derivative couplings become
\begin{equation}
  \label{eq:q4all}
 \begin{split}
  \mathcal{L}_4^{(4)} =
  & -\left[A^{1234}+\Sigma^{1234}\right]s^{1}_2\nabla_{\mu}s^{2}_2\nabla\cdot\nabla(s^{3}_3\nabla^{\mu}s^{4}_3)\\
  & -\left[A^{3412}+\Sigma^{3412}\right]s^{1}_3\nabla_{\mu}s^{2}_3\nabla\cdot\nabla(s^{3}_2\nabla^{\mu}s^{4}_2)\\
  & +\left[A^{1324}+\Sigma^{1324}\right]s^{1}_2\nabla_{\mu}s^{2}_3\nabla\cdot\nabla(s^{3}_2\nabla^{\mu}s^{4}_3)\\
  & +\left[A^{2413}+\Sigma^{2413}\right]s^{1}_3\nabla_{\mu}s^{2}_2\nabla\cdot\nabla(s^{3}_3\nabla^{\mu}s^{4}_2)\\
  & +A^{4123}s^{1}_2\nabla_{\mu}s^{2}_3\nabla\cdot\nabla(s^{3}_3\nabla^{\mu}s^{4}_2)\\
  & +A^{3214}s^{1}_3\nabla_{\mu}s^{2}_2\nabla\cdot\nabla(s^{3}_2\nabla^{\mu}s^{4}_3)\\
 \end{split}
\end{equation}

To rewrite these terms, we first note that on an $AdS_5$ background, we have the following important
formula, obtained by explicitly writing out the $\nabla \cdot \nabla$ derivative in the four-derivative
terms, and use that
\[
[\nabla_\mu, \nabla_\nu] V^\rho = R^\rho_{\;\sigma\mu\nu} V^\sigma,
\]
where $R^\rho_{\;\sigma\mu\nu}$ is the Riemann tensor associated with the Euclidean metric on
$AdS_5$. Applying this formula leads to $\nabla^2 \nabla_\mu s_k^{I_k} = \nabla_\mu (\nabla^2 - 4)
s_k^{I_k}$. Using this result we find
\begin{equation} \begin{split} s_{k_1}^1 \nabla_\mu s_{k_2}^2
  &\nabla\cdot\nabla (s_{k_3}^3 \nabla^\mu s_{k_4}^3) = \\&(m_{k_3}^2 + m_{k_4}^2-4) s_{k_1}^1
  \nabla_\mu s_{k_2}^2
  s_{k_3}^3 \nabla^\mu s_{k_4}^4 + 2 s_{k_1}^1 \nabla_\mu s_{k_2}^2 \nabla_\nu s_{k_3}^3 \nabla^\nu
  \nabla^\mu s_{k_4}^4,
\end{split} \label{eq:quarticads} \end{equation}
where in the equations of motion for the $s$-fields on an $AdS$-background are
used.\footnote{The equation of motion we use is $(\nabla^2-m_k^2) s_k = 0$, that is, we do not
  include the correction terms, since they do not give contributions to the four-point function.
  This is because these correction terms include the fields on $AdS_5$, hence this would lead to
  terms with five fields or more.}

If we now use formula~\eqref{eq:quarticads} on each line in eq.\eqref{eq:q4all}, it is easy to see,
after relabeling the summation indices, that the four-derivative terms in the first four lines
cancel each other. For the last two lines, recall that $A^{1234}$ is anti-symmetric under the
interchange of $3$ and $4$. But $s_2^1 \nabla_\mu s_3^2 \nabla_\nu s_3^3 \nabla^\mu \nabla^\nu s_3^4$
is symmetric under $2 \leftrightarrow 3$, hence $A^{4123} s_2^1 \nabla_\mu s_3^2 \nabla_\nu s_3^3
\nabla^\mu \nabla^\nu s_3^4$ must vanish. A similar argument holds for the last line, hence we
conclude that \emph{all four-derivative terms vanish}.

If we sum and relabel the remaining terms after applying equation~\eqref{eq:quarticads}, we see that
the four-derivative couplings give a contribution to the two-derivative couplings, given by
\begin{equation}
  \begin{split}
    \mathcal{L}^{(4)}_{4} = \Sigma^{1234} (m_2^2&+m_3^2-4) (-2 s_2^1 \nabla_\mu s_2^2 s_3^3 \nabla^\mu
  s_3^4 \\
  &+ s_2^1 \nabla_\mu s_3^3 s_2^2 \nabla^\mu s_3^4 + s_3^3 \nabla_\mu s_2^1 s_3^4 \nabla^\mu s_2^2),
\end{split}
  \label{eq:four2two}
\end{equation}
since the terms $A^{1234}$ times an expression symmetric in $3 \leftrightarrow 4$ vanish.

We can simplify this even further. Notice that by a partial
integration\footnote{The boundary terms do not contribute to the
four-point function, see~\cite{Arutyunov:2000im}.} we have for a
general tensor $\chi^{1234}$
\begin{equation}
 \begin{split}
   \chi^{1234}s^1_2 \nabla_{\mu}s^{3}_3 s^2_2 \nabla^{\mu}s^{4}_3  = -(\chi^{2134} + \chi^{1234})
   s_2^1 \nabla_\mu s_2^2 s_3^3 \nabla^\mu s_3^4 - m_3^2 s_2^1 s_2^2 s_3^3 s_3^4,
  \end{split}
  \label{eq:partialint}
\end{equation}
where the linearized equation of motion is used again. If we use
this in eq.\eqref{eq:four2two} and the symmetry properties
of the tensor $\Sigma^{1234}$, it reduces to
\begin{equation}
    \mathcal{L}^{(4)}_{4} = \Sigma^{1234} (m_2^2+m_3^2-4) (-6 s_2^1 \nabla_\mu s_2^2 s_3^3 \nabla^\mu
  s_3^4 - (m_2^2+m_3^2) s_2^1 s_2^2 s_3^3 s_3^4 ).
  \label{eq:four2twozero}
\end{equation}
This completes the calculation of the four-derivative couplings.

\subsection{Two-derivative couplings}
Proceeding in the same manner as before, we find for the two-derivative couplings
\begin{align*}
  \mathcal{L}_4^{(2)} & = - \frac{1}{165150720}[337869022C^{1234}+296581342C^{1243}-24924719\delta^{12}\delta^{34}\\
  & \quad \quad \quad \quad \quad \quad +861160876S^{1234}+62232811(\Upsilon^{1234}+\Upsilon^{1243})]s^{1}_2\nabla_{\mu}s^{2}_2s^{3}_3\nabla^{\mu}s^{4}_3\\
  &
  \quad -\frac{1}{660602880}[62281822C^{1234}+529982734C^{1243}-24999563\delta^{12}\delta^{34}\\
  & \quad \quad \quad \quad \quad \quad \quad +901550428S^{1234}-56591825\Upsilon^{1234}\\
  & \quad \quad \quad \quad \quad \quad \quad +180608383\Upsilon^{1243}]
  (s_3^3 \nabla_\mu s_2^1 s_3^4 \nabla^\mu s_3^2 + s_2^1 \nabla_\mu s_3^3 s_2^2 \nabla^\mu s_3^4)
\end{align*}
If we now use eq.\eqref{eq:partialint} again, we can rewrite this as
\begin{equation}
 \begin{split}
  \mathcal{L}_4^{(2)} & = - \frac{1}{655360}[165622C^{1234}+1782C^{1243}+297\delta^{12}\delta^{34}\\
  & \quad \quad -160276S^{1234}+891(\Upsilon^{1234}+\Upsilon^{1243})]s^{1}_2\nabla_{\mu}s^{2}_2s^{3}_3\nabla^{\mu}s^{4}_3\\
  & +\frac{(m^2_3+m^2_2)}{660602880}[62281822C^{1234} + 529982734C^{1243}-24999563\delta^{12}\delta^{34}\\
  & \quad \quad +901550428S^{1234}-56591825
  \Upsilon^{1234}+180608383\Upsilon^{1243}]s^{1}_2s^{2}_2s^{3}_3s^{4}_3.
 \end{split}
\end{equation}

\subsection{Non-derivative couplings}
The non-derivative terms contribute, after using the symmetry under $1 \leftrightarrow 2$ and $3
\leftrightarrow 4$, by
\[
  \begin{split}
    \mathcal{L}^{(0)}_4 =  \frac{1}{94371840}  \big[ 911368268 & C^{1234} + 1079096380 S^{1234}
    - 60339107 \delta_2^{12}\delta_3^{34} \\ & + 18147614 \Upsilon^{1234} \big]
    s_2^1 s_2^2 s_3^3 s_3^4.
\end{split}
\]
A remarkable thing happens when we now add all quartic terms: all the ``ugly'' coefficients add up
to a very simple result
\begin{equation}
  \begin{split}
  \mathcal{L}_4 = -\frac{1}{4}&\left( C^{1234} - S^{1234} \right) s_2^1 \nabla_\mu s_2^2 s_3^3 \nabla^\mu
  s_3^4 \\ &+ \frac{3}{8}\left( 9 C^{1234} + 5 S^{1234} - \delta_2^{12} \delta_3^{34} - 3 \Upsilon^{1234}
  \right) s_2^1 s_2^2 s_3^3 s_3^4.
  \end{split}
  \label{eq:quarticfinal}
\end{equation}
This also happened in all other cases treated so far, where four operators of equal weight
($k=2,3,4$) were
considered~\cite{Arutyunov:2000py,Arutyunov:2002fh,Arutyunov:2003ae}.\footnote{After the completion
of this calculation, we learned from~\cite{Arutyunov:2000im}, that
the same result was proved that these
four-derivative couplings have to vanish for the AdS/CFT correspondence to be consistent. This is a
so-called sub-subextremal case, where $k_1 = k_2+k_3+k_4-4$. If this coupling were non-zero, the
associated contact diagram would lead to divergences. The calculation done there is in the same
spirit as this one.}

\section{$D$-functions}\label{ap:dfunc}
In the evaluation of the graphs contributing to the four-point functions, an important r\^{o}le is
played by the so-called \emph{$D$-functions}. These $D$-functions correspond to a quartic
interaction of scalar fields~\cite{D'Hoker:1999pj}. The $D$-functions related to $\AdS$ are defined
by
\begin{equation}
  D_{\Delta_1 \Delta_2 \Delta_3 \Delta_4}(x_1, x_2, x_3, x_4) = \int \frac{d^5z}{z_0^5}
  K_{\Delta_1}(z, x_1)K_{\Delta_2}(z, x_2)K_{\Delta_3}(z, x_3)K_{\Delta_4}(z, x_4),
  \label{eq:dfunction}
\end{equation}
where $K_\Delta$ is the bulk-to-boundary propagator for scalar fields, defined by
\[
K_\Delta(z,\vec{x}) = \left( \frac{z_0}{z_0^2 + (\vec{z}-\vec{x})^2)} \right)^\Delta.
\]

It is possible to express the $D$-functions in terms of the conformal cross-ratios $s$ and $t$.
Introducing the notation $\Db$, we have based on the conformal symmetries
\begin{equation}
  \frac{\prod_{i=1}^4 \Gamma(\Delta_i)}{\Gamma(\Sigma-2)} \frac{2}{\pi^{2}}
  D_{\Delta_1 \Delta_2 \Delta_3 \Delta_4} = \frac{(x_{14}^2)^{\Sigma-\Delta_1-\Delta_4}
  (x_{34}^2)^{\Sigma-\Delta_3-\Delta_4}}{(x_{13}^2)^{\Sigma-\Delta_4}(x_{24}^2)^{\Delta_2}}
  \Db_{\Delta_1 \Delta_2 \Delta_3 \Delta_4}(s,t),
  \label{eq:dbar}
\end{equation}
where $\Sigma = \frac{1}{2} \sum_{i=1}^4 \Delta_i$. For $\Delta_i = 1$, this expression becomes
\begin{equation}
  \Db_{1111}(s,t) = \Phi(s,t),
  \label{eq:boxint}
\end{equation}
where $\Phi(s,t)$ is the one-loop (box) integral as a function of
the conformal cross-ratios~\cite{Usyukina:1992jd}.  There are a number of relations for
these $\Db$-functions, which can be used to simplify expressions and
check crossing symmetry. We do not need them here, but see
Ref.~\cite{Arutyunov:2002fh} for a list.

By considering relations for derivatives of the originial $D$-functions, it is possible to derive
the following relations for the $\Db$-functions~\cite{Arutyunov:2002fh}
\begin{equation}
  \begin{split}
    \Db_{\Delta_1+1\Delta_2+1 \Delta_3 \Delta_4} & = - \p_s \Db_{\Delta_1\Delta_2\Delta_3\Delta_4} \\
    \Db_{\Delta_1\Delta_2\Delta_3+1\Delta_4+1} & = (\Delta_3+\Delta_4-\Sigma - s \p_s) \Db_{\Delta_1\Delta_2\Delta_3\Delta_4} \\
    \Db_{\Delta_1\Delta_2+1 \Delta_3+1 \Delta_4} & = -\p_t \Db_{\Delta_1\Delta_2\Delta_3\Delta_4} \\
    \Db_{\Delta_1+1\Delta_2 \Delta_3 \Delta_4+1} & = (\Delta_1+\Delta_4-\Sigma - t\p_t) \Db_{\Delta_1\Delta_2\Delta_3\Delta_4} \\
    \Db_{\Delta_1\Delta_2+1 \Delta_3 \Delta_4+1} & = (\Delta_2+s \p_s + t \p_t) \Db_{\Delta_1\Delta_2\Delta_3\Delta_4} \\
    \Db_{\Delta_1+1\Delta_2 \Delta_3+1 \Delta_4} & = (\Sigma-\Delta_4+ s\p_s + t \p_t)
    \Db_{\Delta_1\Delta_2\Delta_3\Delta_4}.
  \end{split}
  \label{eq:dbarrel}
\end{equation}
Starting with~\eqref{eq:boxint} and subsequently applying these relations, it is possible to assign
to each $D$-function a differential operator $\D$\index{D operator@$\D$ operator}, such that
\begin{equation}
  \Db_{\Delta_1 \Delta_2 \Delta_3 \Delta_4}(s,t) = \D_{\Delta_1 \Delta_2 \Delta_3 \Delta_4} \Phi(s,t),
  \label{eq:dbarbox}
\end{equation}
as long as each $\Delta_1$ is an integer, and their sum is even. Note that there is some
arbitrariness in defining $\D$, because there are different combinations of the relations
in eqs.\eqref{eq:dbarrel} that one can use to find a particular differential operator.

The action of the partial derivatives on $\Phi(s,t)$ is known~\cite{Eden:2000bk}, they are given by
\begin{equation}
  \begin{split}
\p_s \Phi(s,t) &= \frac{1}{\lambda^2} \left( \Phi(s,t)(1-s+t) + 2 \ln s - \frac{s+t-1}{s} \ln t \right) \\
\p_t \Phi(s,t) &= \frac{1}{\lambda^2} \left( \Phi(s,t)(1-t+s) + 2 \ln t - \frac{s+t-1}{t} \ln s
\right),
  \end{split}
  \label{eq:boxpartial}
\end{equation}
where $\lambda = \sqrt{(1-s-t)^2 - 4 s t}$. Using this together with
eq.~\eqref{eq:dbarbox}, it is possible to express each combination of
$D$-functions into an expression involving only $\Phi(s,t)$.  This is
the method we will use to check the partial non-renormalization of the
four-point function.

\section{Exchange Graphs}\label{sec:exchange}
In this section we generalize the method of~\cite{D'Hoker:1999ni}
and Appendix E of~\cite{Arutyunov:2002fh} to calculate the
$z$-integrals in exchange diagrams. Our results include couplings
to scalar fields of different mass. Together with the previous
results cited above, all exchange diagrams contributing to
\emph{arbitrary} four-point functions of $\frac{1}{2}$-BPS
operators can be calculated.

We briefly review the method used: First, conformal symmetry is
used to bring the $z$-integral into a simpler form. The basic idea
is then to use the wave equation for the propagator in the
$z$-integral. Based on conformal invariance an ansatz for the
$z$-integral is proposed. Then, the wave equation is applied to
the integrand of the $z$-integral and on the ansatz. This
leads to a system of differential equations, from which we can
solve for the $z$-integral.

\subsection{Vector Exchange} \label{VEG}
Here we generalize the calculation of the vector $z$-integral done in \cite{D'Hoker:1999ni} to
include the case when the vector field couple to scalar fields of different mass. The $z$-integral
is given by
$$
A_{\mu}(\omega,\vec{x}_1,\vec{x}_3)=\int\frac{d^{d+1}z}{z_0^{d+1}}G_{\mu\nu'}(\omega,z)g^{\nu'\rho'}(z)K_{\Delta_1}(z,\vec{x}_1)\frac{\overleftrightarrow{\partial}}{\partial
z^{\rho'}}K_{\Delta_3}(z,\vec{x}_3),
$$
where $K_{\Delta}(z,\vec{x})$ is the canonically normalized bulk-to-boundary propagator for a scalar
field $s_{\Delta}$.

The gauge boson propagator $G_{\mu\nu'}(\omega,z)$ with mass $M$ in $AdS_5$ satisfies the defining
wave equation:
\begin{equation}
\label{vpe}
-\nabla^{\mu}\nabla_{[\mu}G_{\rho]\nu'}+M^2G_{\rho\nu'}=g_{\rho\nu'}\delta(\omega,z)+\partial_{\rho}\partial_{\nu'}\Lambda(u),
\end{equation}
where $[..]$ denotes anti-symmetrization.  We can drop the gauge
term $\Lambda(u)$, since this is only necessary for couplings to the
massless gauge boson. However, the gauge boson couples only to
conserved currents and we refer to~\cite{D'Hoker:1999ni} for this
case.

Using
$A_{\mu\nu}(\omega,\vec{x}_1,\vec{x}_3)=A_{\mu\nu}(\omega-\vec{x}_1,0,\vec{x}_{13})$,
where $\vec{x}_{13} \equiv \vec{x}_3-\vec{x}_1$, and performing the conformal
inversion $\omega'_{\mu}=\omega_{\mu}/\omega^2$, $x'_{\mu}=x_{\mu}/x^2$ the
vector $z$
integral takes the following form
\begin{equation}
\label{vzi}
 A_{\mu}(\omega,\vec{x}_1,\vec{x}_3)=|\vec{x}_{13}|^{-2\Delta_3}\frac{1}{\omega^2}J_{\mu\nu}(\omega)I_{\mu}(\omega'-\vec{x}'_{13}),
\end{equation}
where $J_{\mu\nu}(\omega)=\delta_{\mu\nu}-\frac{2}{\omega^2}\omega_{\mu}\omega_{\nu}$, and
\begin{equation}
\label{vi1}
 I_{\mu}(\omega)=\int
 \frac{d^{d+1}z}{z^{d+1}_0}{G_{\mu}}^{\nu'}(\omega,z)z_0^{\Delta_1}\frac{\overleftrightarrow{\partial}}{\partial
 z^{\nu'}}\left(\frac{z_0}{z^2}\right)^{\Delta_3}.
\end{equation}

We write the following ansatz, with $\Delta_{13} = \Delta_1 - \Delta_3$,
\begin{equation}
\label{vi2}
I_{\mu}(\omega)=\omega_0^{\Delta_{13}}\frac{\omega_{\mu}}{\omega^2}f(t)+\omega_0^{\Delta_{13}}\frac{\delta_{\mu
0}}{\omega_0}h(t),
\end{equation}
where $t = \omega_0^2/\omega^2$.

To find the unknown scalar functions $f(t)$ and $h(t)$ we equate eqs.\eqref{vi1} and \eqref{vi2} and apply the
differential operator in eq.\eqref{vpe} to both sides. For eq.\eqref{vi1} we obtain
 \begin{equation}
\label{vde}
-\nabla^{\mu}\nabla_{[\mu}I_{\rho]}+M^2I_{\rho}=-\omega_0^{\Delta_{13}}\frac{\omega_{\rho}}{\omega^2}2\Delta_{3}t^{\Delta_{3}}-\omega_0^{\Delta_{13}}\frac{\delta_{\rho0}}{\omega_0}\Delta_{13}t^{\Delta_{3}}.
\end{equation}
For eq.\eqref{vi2} the Maxwell term becomes
\begin{equation}\label{Mt}
 \begin{split}
  -\nabla^{\mu}\nabla_{[\mu}I_{\rho]} & =\omega^2\delta_{\mu\lambda}\partial_{\lambda}T_{\rho\mu}+\omega^2\delta_{\mu\lambda}\Gamma_{\rho\lambda}^{\kappa}T_{\kappa\mu}+\omega^2\delta_{\mu\lambda}\Gamma_{\lambda\mu}^{\kappa}T_{\rho\kappa},
 \end{split}
\end{equation}
where we defined
\begin{equation}\label{T}
 \begin{split}
  T_{\rho\mu} & :=\partial_{[\mu}I_{\rho]}=(\omega_{\rho}\frac{\delta_{\mu0}}{\omega_0}-\omega_{\mu}\frac{\delta_{\rho0}}{\omega_0})\left(\frac{2\omega_0^{\Delta_{13}+2}}{(\omega^2)^2}\left(f'+h'\right)+\Delta_{13}\frac{\omega_0^{\Delta_{13}}}{\omega^2}f(t)\right).
 \end{split}
\end{equation}
We omit the tedious but straightforward calculation of these terms.

Now, substituting the result in the LHS of equation (\ref{vde}), we find the equation in terms of
$f$ and $h$.  To solve it, we equate the corresponding contributions of the tensor structures
$\frac{\delta_{\rho0}}{\omega_0}$ and $\frac{\omega_{\rho}}{\omega^2}$ to the LHS and RHS, and
obtain the following coupled system of inhomogeneous differential equations:
\begin{align}
\begin{split}
2\Delta_{13}&t^2(f'+h')+\Delta_{13}^2tf+M^2h=-\Delta_{13}t^{\Delta_{3}}, \\
\end{split}\label{ide1} \\
 \begin{split}
   4t^2&(t-1)\left(f''+h''\right)+2t[t(4+\Delta_{13})+(d-4-2\Delta_{13})]f'\\
   &+2t[4t+d-4-\Delta_{13}]h'+[\Delta_{13}(2t+d-2-\Delta_{13})+M^2]f=-2\Delta_{3}t^{\Delta_{3}}.\\
 \end{split}\label{ide2}
\end{align}
To solve the differential equations we assume power series expansions
\begin{equation}
 \label{series}
f(t)=\sum_{k}a_kt^k, \quad\quad h(t)=\sum_{k}b_kt^k,
\end{equation}
with $k_{min}\leq k\leq k_{max}$.

Substituting eq.\eqref{series} in eqs.\eqref{ide1} and \eqref{ide2}; and intersecting the equations we find $b_k$ and $b_{k+1}$ in terms of $a_k$ and $a_{k+1}$:
\begin{equation}
  \begin{split}
    & b_k= -\frac{2k+\Delta_{13}}{2k}a_k+\frac{M^2 [2(-2 k-4+d)(k+1)+\Delta_{13}(4k-6+d-\Delta_{13})+M^2]}{4(k+1)k[\Delta_{13}(-2k-4+d-\Delta_{13})-M^2]}a_{k+1}\\
    & +\frac{\Delta_{13}(2\Delta_3+2-d+\Delta_{13})-M^2}{2(\Delta_{13}-1)[(-2\Delta_3-2+d-\Delta_{13})-M^2]}\delta_{k,\Delta_3-1},\\
    & b_{k+1}= -\frac{\Delta_{13}[2(k+1)(-2k-4+d)+\Delta_{13}(-4k-6+d-\Delta_{13})+M^2]}{2(k+1)[\Delta_{13}(-2k-4+d-\Delta_{13})-M^2]}a_{k+1}.\\
  \end{split}
\end{equation}
The solution is found if we substitute the second equation in the first one:
\begin{equation}
\begin{split}
 & a_k=0 \qquad \text{for} \qquad k\geq \Delta_3 \\
 & a_{\Delta_3-1}=\frac{\Delta_{13}(d-2\Delta_3)-\Delta_{13}^2-M^2}{2M^2(\Delta_3-1+\Delta_{13})},\\
 & a_k=\frac{[\Delta_{13}^2+\Delta_{13}(2k+2-d)][2(k+1)(2k+4-d)+\Delta_{13}(4k+6-d)+\Delta_{13}^2-M^2]}{4(k+1)(k+\Delta_{13})[\Delta_{13}(2k+4-d)+\Delta^2+M^2]}a_{k+1}
\end{split}
\end{equation}
From this we see that the series terminates at
$$
 0\leq
 k_{min}=\frac{d-2-2\Delta_{13}}{4}+\frac{1}{4}\sqrt{(d-2)^2+4M^2}\leq
 k_{max}=\Delta_{3}-1,
$$
provided that $k_{max}-k_{min}$ is an integer and $\geq0$.

One can show using Table III  in \cite{Kim:1985ez}, where
$M^2=l^2-1$ with $l\in \mathbb{Z}^+$, that the terminating condition
is always satisfied by Type IIB supergravity compactified in
$AdS_5\times S_5$ due to the $SO(6)$ selection rules
\cite{group}.\footnote{It is worth to recall that the marginal case,
when the equality holds and the series doesn't terminates, is also
allowed by the $SO(6)$ selection rules. We consider here only the
terminating case. Note also that the case $M^2=0$ requires
$\Delta_1+\Delta_3-d \in \mathbb{N}$ for terminating of the series.}

Finally, to recover the vector $z$-integral in terms of the original
coordinates, we must transform the coordinates back. This amounts to
\begin{equation}\label{ttransf}
\begin{split}
\omega_0' & \rightarrow
 \frac{\omega_0}{\omega_0^2+(\vec{\omega}-\vec{x}_1)^2},\\
 t=\frac{{\omega'_0}^2}{(\omega'-\vec{x}'_{31})^2} & \rightarrow
 q=\vec{x}_{31}^2\frac{\omega_0}{\omega_0^2+(\vec{\omega}-\vec{x}_1)^2}\frac{\omega_0}{\omega_0^2+(\vec{\omega}-\vec{x}_3)^2},\\
 \frac{1}{\omega^2}J_{\mu\lambda}(\omega)\frac{(\omega'-\vec{x}'_{31})_{\lambda}}{(\omega'-\vec{x}'_{31})^2}
 & \rightarrow
 Q_{\mu}:=\frac{(\omega-\vec{x}_3)_{\mu}}{(\omega-\vec{x}_3)^2}-\frac{(\omega-\vec{x}_1)_{\mu}}{(\omega-\vec{x}_1)^2}\\
 \frac{1}{\omega^2}J_{\mu\lambda}(\omega)\frac{\delta_{\lambda0}}{\omega'_0} & \rightarrow
 R_{\mu}:=\frac{\delta_{\mu0}}{\omega_0}-2\frac{(\omega-\vec{x}_1)_{\mu}}{(\omega-\vec{x}_1)^2}.\\
\end{split}
\end{equation}

In the case we need in this paper, $\Delta_1=3$, $\Delta_3=2$ and $M^2=3$, we find
\begin{equation}
  \begin{split}
    A_{\mu}(\omega,\vec{x}_1,\vec{x}_3)= & \frac{1}{\vec{x}_{31}^{2}}\left[-\frac{1}{12}K_1(\omega,\vec{x}_3)\nabla_{\mu}K_2(\omega,\vec{x}_1)+\frac{1}{6}K_2(\omega,\vec{x}_1)\nabla_{\mu}K_1(\omega,\vec{x}_3)\right].\\
  \end{split}
\end{equation}
The $\omega$-integral can then be calculated straightforwardly.

\subsection{Symmetric Tensor Exchange} \label{TEG}
Here we extend the computation in \cite{Arutyunov:2002fh} using the method of \cite{D'Hoker:1999ni}
for the massive symmetric tensor $z$-integral when the tensor field is coupled to scalar fields of
different mass. In this case the stress energy tensor $T_{\mu\nu}$ is not covariantly conserved and
has the form
\begin{equation}
T_{\mu\nu}=\frac{1}{2}\nabla_{(\mu}s_{\Delta_1}\nabla_{\nu)}s_{\Delta_3}-\frac{1}{2}g_{\mu\nu}\left(\nabla^{\rho}s_{\Delta_1}\nabla_{\rho}s_{\Delta_3}+\frac{1}{2}(m_{\Delta_1}^2+m_{\Delta_3}^2-f)s_{\Delta_1}s_{\Delta_3}\right),
\end{equation}
where $s_{\Delta}$ denotes a scalar field of mass squared
$m_{\Delta}^2=\Delta(\Delta-4)$, and $f$ is the mass squared of the tensor.

The $z$-integral describing the exchange of a massive symmetric tensor is given by
\begin{equation}
 \label{int}
A_{\mu\nu}(\omega,\vec{x}_1,\vec{x}_3):=\int\frac{d^{d+1}z}{z_0^{d+1}}G_{\mu\nu\mu'\nu'}(\omega,z)T^{\mu'\nu'}(z,\vec{x}_1,\vec{x}_3).
\end{equation}
The tensor $T^{\mu\nu}(z,\vec{x}_1,\vec{x}_3)$ has the form
\begin{equation}
 \label{tensstres}
\begin{split}
T^{\mu\nu}&(z,\vec{x}_1,\vec{x}_3)=  \frac{1}{2}\nabla^{(\mu}K_{\Delta_1}(z,\vec{x}_1)\nabla^{\nu)}K_{\Delta_3}(z,\vec{x}_3)
-
\frac{1}{2}g^{\mu\nu}\left[\nabla_{\rho}K_{\Delta_1}(z,\vec{x}_1)\nabla^{\rho}K_{\Delta_3}(z,\vec{x}_3)
\right.
\\
&\left.+\frac{1}{2}\left(m_{\Delta_1}^2+m_{\Delta_3}^2-f\right)K_{\Delta_1}(z,\vec{x}_1)K_{\Delta_3}(z,\vec{x}_3)\right],
\end{split}
\end{equation}
where $(..)$ denotes symmetrization.

The Ricci form of the wave equation for the bulk-to-bulk propagator $G_{\mu\nu\mu'\nu'}(\omega,z)$ for the massive symmetric tensor field  is
\begin{equation}
\label{gravprop}
 \begin{split}
 {W_{\mu\nu}}^{\lambda\rho}[G_{\lambda\rho\mu'\nu'}] &
 =\left(g_{\mu\mu'}g_{\nu\nu'}+g_{\mu\nu'}g_{\nu\mu'}-\frac{2}{d-1}g_{\mu\nu}g_{\mu'\nu'}\right)\delta(\omega,z).
  \end{split}
\end{equation}
Pure gauge terms are omitted, because they are not needed in the case of massive tensors. The
graviton only couples to conserved currents, and we refer
to Appendix E of~\cite{Arutyunov:2002fh} for this case.

To solve the tensor $z$-integral we use again
$A_{\mu\nu}(\omega,\vec{x}_1,\vec{x}_3)=A_{\mu\nu}(\omega-\vec{x}_1,0,\vec{x}_{13})$
and perform the conformal inversion
$\omega'_{\mu}=\omega_{\mu}/\omega^2$,
$\vec{x}'_{\mu}=\vec{x}_{\mu}/x^2$ on eq.\eqref{int} to obtain
\begin{equation}
\label{tzi}
 A_{\mu\nu}(\omega,\vec{x}_1,\vec{x}_3)=|\vec{x}_{13}|^{-2\Delta_3}\frac{1}{(\omega^2)^2}J_{\mu\lambda}(\omega)J_{\nu\rho}(\omega)I_{\lambda\rho}(\omega'-\vec{x}'_{13}),
\end{equation}
where
\begin{equation}
\label{ti1}
 \begin{split}
  I_{\mu\nu}(\omega)=\int
  \frac{d^{d+1}z}{z^{d+1}_0} &
  {G_{\mu\nu}}^{\mu'\nu'}(\omega,z)\left\{\nabla_{(\mu'}z_0^{\Delta_1}\nabla_{\nu')}\left(\frac{z_0}{z^2}\right)^{\Delta_3}\right.\\
 & \left.
 -g_{\mu'\nu'}\left[\nabla_{\rho'}z_0^{\Delta_1}\nabla^{\rho'}\left(\frac{z_0}{z^2}\right)^{\Delta_3}+\frac{1}{2}(m_1^2+m_2^2-f)z_0^{\Delta_1}\left(\frac{z_0}{z^2}\right)^{\Delta_3}\right]\right\}.\\
 \end{split}
\end{equation}
We write the following ansatz
\begin{equation}
\label{ti2}
 \begin{split}
  I_{\mu\nu}(\omega) &
  =\omega_0^{\Delta_{13}}g_{\mu\nu}h(t)+\omega_0^{\Delta_{13}}P_{\mu}P_{\nu}\phi(t)+\omega_0^{\Delta_{13}}\nabla_{\mu}\nabla_{\nu}X(t)+\omega_0^{\Delta_{13}}\nabla_{(\mu}\left(P_{\nu)}Y(t)\right)\\
  & =\omega_0^{\Delta_{13}}\tilde{I}_{\mu\nu}(\omega),
 \end{split}
\end{equation}
where $P_{\mu}:=\delta_{\mu0}/\omega_0$; $h(t)$, $\phi(t)$,
$X(t)$, $Y(t)$ are four unknown scalar functions, and
$\tilde{I}_{\mu\nu}(\omega)$ is the ansatz used in \cite{Arutyunov:2002fh} for the case $\Delta_1=\Delta_3$.
To find the functions we first have to equate eqs.~(\ref{ti1}) and
(\ref{ti2}) and apply the modified Ricci operator on both sides. For
eq.\eqref{ti2} we first note that
\begin{align*}
 \nabla_{\rho}\nabla_{\sigma}I_{\mu\nu} & =\nabla_{\rho}\nabla_{\sigma}\left(\omega_0^{\Delta_{13}}\tilde{I}_{\mu\nu}\right)\\
 & =\omega_0^{\Delta_{13}}\left(\nabla_{\rho}\nabla_{\sigma}\tilde{I}_{\mu\nu}\right)+\left(\nabla_{(\rho}\omega_0^{\Delta_{13}}\right)\left(\nabla_{\sigma)}\tilde{I}_{\mu\nu}\right)+\left(\nabla_{\rho}\nabla_{\sigma}\omega_0^{\Delta_{13}}\right)\tilde{I}_{\mu\nu}.\\
\end{align*}
This identity allows us to write
\begin{equation}
\label{R}
 {W_{\mu\nu}}^{\lambda\rho}[I_{\lambda\rho}]=\omega_0^{\Delta_{13}}\left({W_{\mu\nu}}^{\lambda\rho}[\tilde{I}_{\rho\sigma}]\right)+H_{\mu\nu}+N_{\mu\nu},
\end{equation}
 where
\begin{align*}
 H_{\mu\nu} & =\left[-\left(\nabla^2\omega_0^{\Delta_{13}}\right)\tilde{I}_{\rho\sigma}-\left(\nabla_{\mu}\nabla_{\nu}\omega_0^{\Delta_{13}}\right){I^{\sigma}}_{\sigma}+\left(\nabla_{\mu}\nabla^{\sigma}\omega_0^{\Delta_{13}}\right)I_{\sigma\nu}+\left(\nabla_{\nu}\nabla^{\sigma}\omega_0^{\Delta_{13}}\right)I_{\mu\sigma}\right],\\
 N_{\mu\nu} & =\left[-2\left(\nabla_{\sigma}\omega_0^{\Delta_{13}}\right)\left(\nabla^{\sigma}\tilde{I}_{\rho\sigma}\right)-\left(\nabla_{(\mu}\omega_0^{\Delta_{13}}\right)\left(\nabla_{\nu)}{I^{\sigma}}_{\sigma}\right)\right.\\
 & \quad \left. +g^{\sigma\kappa}\left(\nabla_{(\mu}\omega_0^{\Delta_{13}}\right)\left(\nabla_{\kappa)}I_{\sigma\nu}\right)+g^{\sigma\kappa}\left(\nabla_{(\nu}\omega_0^{\Delta_{13}}\right)\left(\nabla_{\kappa)}I_{\mu\sigma}\right)\right].\\
\end{align*}
The first term in eq.\eqref{R} was calculated in Appendix
E of \cite{Arutyunov:2002fh}. For the $H_{\mu\nu}$ and $N_{\mu\nu}$ we obtain the
following formulae
\begin{equation}
 \label{extra}
\begin{split}
 H_{\mu\nu}=\omega_0 & ^{\Delta_{13}}\Delta_{13}\left\{g_{\mu\nu}\left[(2d-1-\Delta_{13})h+\phi+4t^2(1-t)X''+2t(2-2t-d)X'\right.\right.\\
 & \quad \quad \quad \left.+4t(1-t)Y'+[4(-d+1)+2\Delta_{13}]Y\right]\\
 & +P_{\mu}P_{\nu}\left[(\Delta_{13}+1)(-d+1)h+(d-1)\phi+(\Delta_{13}+1)4t^2(1-t)X''\right.\\
 & \quad \quad \quad \left.+(\Delta_{13}+1)2t(2-4t+d)X'+4t(d-1)Y'+(4+2\Delta_{13})(d-1)Y\right]\\
 & +(d-2-\Delta_{13})\nabla_{\mu}\nabla_{\nu}X\\
 & \left.+P_{(\mu}\frac{\omega_{\nu)}}{\omega^2}\left[(\Delta_{13}+1)4t^2(1-t)X''+(\Delta_{13}+1)2t(4t-3)X'+2t(-d+1)Y'\right]\right\},\\
 N_{\mu\nu}=\omega_0 & ^{\Delta_{13}}\Delta_{13}\left\{g_{\mu\nu}\left[4t(t-1)h'-2\phi+4t(t-1)X'+4t(1-t)Y'-4Y\right]\right.\\
 & +P_{\mu}P_{\nu}\left[4t(-d+1)h'+2(-d+1)\phi+4t(-d+1)X'\right.\\
 & \left. \quad \quad \quad +8t^2(t-1)Y''+4t(4t-3)Y'+4(-d+1)Y\right]\\
 & +2\nabla_{\mu}\nabla_{\nu}Y\\
 & \left.+P_{(\mu}\frac{\omega_{\nu)}}{\omega^2}\left[2t(d-1)h'+2t(d-1)X'+8t^2(1-t)Y''+2t(-d+8-8t)Y'\right]\right\}.\\
\end{split}
\end{equation}

For eq.\eqref{ti1}, due
to the defining wave equation for the propagator of the symmetric
tensor field eq.(\ref{gravprop}), we obtain 
\begin{equation}
\label{tde}
 \begin{split}
  {W_{\mu\nu}}^{\lambda\rho}[I_{\lambda\rho}]= &
\omega_0^{\Delta_{13}}g_{\mu\nu}\frac{2}{d-1}(m_1^2+m_2^2-f)t^{\Delta_3}+\omega_0^{\Delta_{13}}P_{\mu}P_{\nu}4\Delta_1\Delta_3t^{\Delta_3}\\
  & -\omega_0^{\Delta_{13}}P_{(\mu}\frac{\omega_{\nu)}}{\omega^2}4\Delta_1\Delta_3t^{\Delta_3}.
 \end{split}
\end{equation}

From eqs.\eqref{R} and \eqref{extra} the basic eq.  (\ref{tde}) can
be written in terms of the four unknown functions in eq.
(\ref{ti2}). To determine them we first equate the terms involving
$\nabla_{\mu}\nabla_{\nu}$ in both sides: \begin{equation}
\label{dd}\nabla_{\mu}\nabla_{\nu}[-3h-\phi+(d\Delta_{13}-2\Delta_{13}-\Delta_{13}^2+f)X+2\Delta_{13}Y]=0.
\end{equation}

Now we equate the coefficients of the tensor
$P_{(\mu}\frac{\omega_{\nu)}}{\omega^2}$ to get \begin{equation}
 \label{po}
 \begin{split}
  2\Delta_{13} & (d-1)h'+4t(t-1)\phi''+8t\phi'\\
  & +(\Delta_{13}^2+\Delta_{13})4t(t-1)X''+[\Delta_{13}^22(4t-3)+\Delta_{13}2(4t-4+d)]X'\\
  & +\Delta_{13}8t(1-t)Y''+2[-f+\Delta_{13}(8-8t-2d)]Y'=-4\Delta_1\Delta_3t^{\Delta_3-1}.
 \end{split}
\end{equation}
This equation can be integrated to give 
\begin{equation}
\label{point}
 \begin{split}
   4t(&t-1)\phi'+4\phi+(\Delta_{13}^2+\Delta_{13})4t(t-1)X'+2(\Delta_{13}d-2\Delta_{13}-\Delta_{13}^2)X\\
  &
  +\Delta_{13}8t(1-t)Y'+2(-f+4\Delta_{13}-2d\Delta_{13})Y + 2\Delta_{13}(d-1)h=-4\Delta_1t^{\Delta_3}+c_1,
 \end{split}
\end{equation}
where $c_1$ is an integration constant.  Equating the coefficients
of $P_{\mu}P_{\nu}$, we get
\begin{equation*} \label{pp}
 \begin{split}
 4 & t\Delta_{13}(-d+1)h'+(\Delta_{13}^2+\Delta_{13})(-d+1)h+4t^2(1-t)\phi''-8t^2\phi'+[\Delta_{13}(-d+1)+f]\phi\\
 & +(\Delta_{13}^2+\Delta_{13})4t^2(1-t)X''+[\Delta_{13}^22t(2-4t+d)+\Delta_{13}2t(4-4t-d)]X'\\
 & +\Delta_{13}8t^2(t-1)Y''+4t[f+\Delta_{13}(4t-4+d)]Y'+2[f+\Delta_{13}^2(d-1)]Y=4\Delta_1\Delta_3t^{\Delta_3}.
 \end{split}
\end{equation*}
Substituting here eq.(\ref{po}) we find \begin{equation}
 \label{popp}
 \begin{split}
  2t\Delta_{13}(-d+1)h' & +(\Delta_{13}^2+\Delta_{13})(-d+1)h+[f+\Delta_{13}(-d+1)]\phi\\
  & +\Delta_{13}^22t(d-1)X'+2tfY'+2[f+\Delta_{13}^2(d-1)]Y=0.
 \end{split}
\end{equation}
Finally, we equate the coefficients of $g_{\mu\nu}$
\begin{equation}
\label{ge}
 \begin{split}
  4t^2 & (t-1)h''+[4t(t+1)+\Delta_{13}4t(t-1)]h'+\left[\frac{8}{3}(f+3)+2d\Delta_{13}-\Delta_{13}-\Delta_{13}^2\right]h\\
  &
  +4t(t-1)\phi'+\left[-\Delta_{13}+\frac{1}{3}(f+24)\right]\phi\\
  & +\left[\frac{f}{3}+\Delta_{13}\right]4t^2(1-t)X''+\left[-\frac{4f}{3}t(t+1)-2td\Delta_{13}\right]X'\\
  & +\left[\frac{f}{3}+2\Delta_{13}\right]4t(1-t)Y'+\left[-\frac{14f}{3}+2\Delta_{13}^2-4\Delta_{13}d\right]Y=\frac{2}{d-1}(m_1^2+m_2^2-f)t^{\Delta_3}.
 \end{split}
\end{equation}

Eqs.(\ref{dd}), (\ref{point}), (\ref{popp}) and (\ref{ge})
form a system of four differential equations whose solution, regular
as $t\rightarrow0$ and $t\rightarrow1$, determines the tensor $z$-integral. We solve them for our specific case: $d=4$, $\Delta_1=3$,
$\Delta_3=2$ and $f=5$; which imply $\Delta_{13}=1$, $m_1^2=-3$ and
$m_3^2=-4$:
\begin{align}
 \nabla_{\mu}\nabla_{\nu}(-3h-\phi+6X+2Y) & =0, \label{eq1}\\
 6h+4t(t-1)\phi'+4\phi+8t(t-1)X'+2X-8t(t-1)Y'-18Y & =-12t^2+c_1, \label{eq2}\\
 6th'+6h-2\phi-6tX'-10tY'-16Y & =0, \label{eq3}\\
 \begin{split} \label{eq4}
  12t^2(t-1)h''+24t^2h'+82h+12t(t-1)\phi'+26\phi \qquad \qquad \qquad & \\
  +32t^2(1-t)X''+[-20t^2-44t]X'+44t(t-1)X-112Y & =-24t^2.
 \end{split}
\end{align}

From eq.(\ref{eq1}) we pick up the trivial solution
\begin{equation}
 \label{h} h=-\frac{1}{3}\phi+2X+\frac{2}{3}Y 
\end{equation}
and substitute it in eq.(\ref{eq3}) to obtain
$$
2t(\phi'-3X'+3Y')+4(\phi-3X+3Y)=0.
$$
This equation can be trivially integrated to give \begin{equation}
 \label{theta}
\phi-3X+3Y=c_2t^{-2}. \end{equation}
 This is regular as $t\rightarrow0$ for $c_2=0$. Substituting $h$ and $\phi$ in terms of $X$
and $Y$ in eq.(\ref{eq2}), we get
$$
20t(t-1)(X'-Y')+20(X-Y)=-12t^2+c_1.
$$
The solution consistent with the asymptotic behavior is given by
\begin{equation}
 \label{X} X=Y-\frac{6}{10}t+\frac{c_1}{20}. 
\end{equation}
Finally, we substitute eqs.(\ref{h}), (\ref{theta}) and (\ref{X}) in eq.(\ref{eq4}) and
find
$$
Y(t)=\frac{18t-3c_1}{40}.
$$
Substituting back, we find the remaining three functions
$$
X(t)=\frac{-6t-c_1}{40}, \qquad \phi(t)=\frac{-36t+3c_1}{20},
\qquad h(t)=\frac{12t-3c_1}{20}.
$$
Upon substitution of the four functions
in eq.\eqref{ti2}, all the terms
proportional to the integration constant $c_1$ cancel and the
tensor integral becomes
$$
I_{\mu\nu}(\omega)=\omega_0\left(\frac{3}{5}tg_{\mu\nu}-\frac{9}{5}tP_{\mu}P_{\nu}-\frac{3}{20}\nabla_{\mu}\nabla_{\nu}t+\frac{9}{20}\nabla_{(\mu}\left(P_{\nu)}t\right)\right).
$$
Working out the derivatives, we find
\begin{equation*} \label{I}
I_{\mu\nu}(\omega)=-\frac{6}{5}\omega_0t\frac{\omega_{\mu}\omega_{\nu}}{(\omega^2)^2}.
\end{equation*}

In terms of the original
coordinates, the $z$-integral describing the exchange by a symmetric tensor field of
mass squared $f=5$ is
\begin{equation*}
A_{\mu\nu}(\omega,\vec{x}_1,\vec{x}_3)=-\frac{1}{\vec{x}_{31}^2}\frac{6}{5}Q_{\mu}Q_{\nu}K_2(\omega,\vec{x}_1)K_1(\omega,\vec{x}_3),
\end{equation*}
where $Q_\mu$ is defined in eq.~\eqref{ttransf}.

\raggedright

\end{document}